\documentclass{PoS}

\usepackage{epsfig,macros_static}

\newcommand{\onecol}[2]{
        \begin{minipage}[t]{#1}{#2\vfill} \end{minipage}
        }

\PoS{PoS(LAT2005)224}

\title{A quenched study of $m_b$ in HQET beyond the leading order}

\ShortTitle{A quenched study of $m_b$ in HQET beyond the leading order}

\author{\epsfxsize=2.5 true cm
        \epsfbox{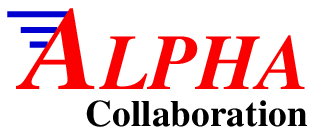} \vspace{-1.8cm} \hfill
{\onecol{2.8cm}{\vspace{-1cm} \it DESY 05-173 \\  SFB/CPP-05-66 \\ HU-EP-05/61}}
}

\author{Michele Della Morte \thanks{
This work has been supported by the Deutsche
Forschungsgemeinschaft (DFG) in
SFB Transregio 9 ``Computational Particle Physics''.}\\

        Institut f\"ur Physik, Humboldt Universit\"at,
        Newtonstr. 15, 12489 Berlin, Germany\\
        E-mail: \email{dellamor@physik.hu-berlin.de}}

\author{\speaker{Nicolas Garron}, Rainer Sommer
        \\
        DESY, 
        Platanenallee 6, 
        15738 Zeuthen, 
        Germany
        \\
        E-mail: \email{Nicolas.Garron@desy.de, Rainer.Sommer@desy.de}}

\author{Mauro Papinutto \\
        NIC/DESY, 
        Platanenallee 6, 
        15738 Zeuthen, 
        Germany
        \\
         E-mail: \email{Mauro.Papinutto@desy.de}}

\abstract{
A non perturbative method to compute 
the mass of the b quark including the $1/m$ term in HQET has been 
presented in~\cite{hqet:Raiproc}. Following this strategy, we find
in the $\overline {\mbox{MS}}$ scheme 
$\m_{\rm b}^{\rm stat}(\m_{\rm b}) = 4.350(64) \,\GeV$ 
for the leading term, and 
$\m_{\rm b}^{(1)}(m_{\rm b}) = -0.049(29) \,\GeV$ 
for the next to leading order correction. 
This method involves several steps,
including the simulation of the relativistic theory in a small 
volume, and of the effective theory in a big volume.
Here we present some numerical details of our calculations.}

\FullConference{XXIIIrd International Symposium on Lattice Field Theory\\

                 25-30 July 2005\\

                 Trinity College, Dublin, Ireland}

\begin{document}
\section{Introduction}
The b quark mass enters in the theoretical predictions of B mesons decay rates, which
provide some of the most precise constraints in the unitarity triangle analysis.
The Particle Data Group~\cite{PDG04} quotes a value
of $\m_{\rm b}(\m_{\rm b})$ between 4.1 and 4.4 GeV in the $\overline {\mbox{MS}}$
scheme, as the result of different determinations 
which include lattice computations.

On the lattice, heavy quarks are treated by using effective theories, in particular
HQET when dealing with heavy-light systems. The b quark mass can then be predicted
using the mass of the B meson as experimental input. 
In this framework, the bare 
parameters must be determined non perturbatively, in order to have a well
defined continuum limit. 
This programme has been completed for the lowest order of the HQET (static theory) 
in the quenched approximation. The result from~\cite{hqet:pap1}, 
$\m_{\rm b}(m_{\rm b})=4.12(11)$ GeV is in 
reasonable agreement with~\cite{deDivitiis:2003iy}, suggesting that the $1/m$ 
corrections to the b quark mass are small. Here we explicitly compute these corrections 
following the strategy proposed in~\cite{hqet:Raiproc}. This clearly represents an 
important test of HQET, which  provides  indications about its ``convergence''.

The method relies on matching QCD and the effective theory in a small volume. 
The observables are then evolved to large
volumes where contact with experiments can be established. Such an evolution
is performed in the effective theory. Different systematics errors 
have to be controlled in the two regimes. For a precise matching in the small volume a very
accurate determination of the relevant renormalization constants in the full theory
is needed. On the other hand, in large volume simulations the main problems usually
concern the separation of the ground state from the excited states.
We present details on the numerical
evaluation of the various ingredients entering the computation of the $1/m$
corrections to the b quark mass. The  increase in the (numerical) effort is moderate
compared to the static case. 
A related investigation is described in~\cite{Negishi:2005cz}. 

\section{Finite volume}

\subsection{Relativistic part}

\begin{figure}[t]
\begin{center}
\leavevmode
\includegraphics[scale=0.3,clip,angle=270]{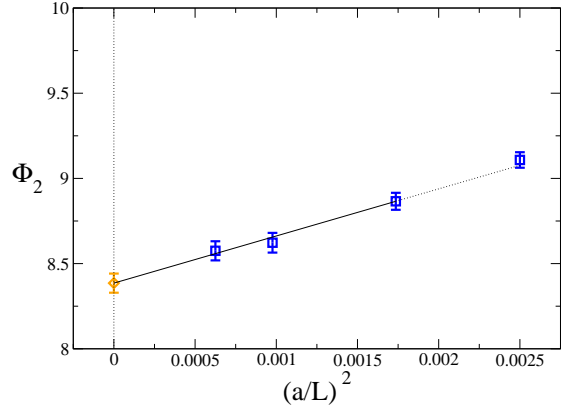}
\end{center}
\vspace{-0.5cm}
\caption[]{\label{phi2} {Continuum extrapolation of $\Phi_2 $, for $L_1M=12$}}
\end{figure}

The observables $\Phi_{1,2}$ are defined as in eqs.~\footnote{
We refer to the equations of~\cite{hqet:Raiproc} with a prefix ``I''.}
(I.2.5), (I.2.6) by ($T=L/2$)
\bes
&\Phi_1(L,M) =& \Phi_1^{\rm rel}(L,M)+\Phi_1^{\rm stat}(L)
\quad \mbox{and} \quad
\Phi_2(L,M)={L \over {2a}} \ln \left( {\foneav(T-a) \over \foneav(T+a)} \right) \quad\\
\mbox{where} \quad
&\Phi_1^{\rm rel}(L,M) =&   \ln \left( {\foneav(\theta,T) \over \foneav(\theta',T)}\right) 
\qquad \mbox{and} \qquad 
\Phi_1^{\rm stat}(L)  = - \ln \left( {\fonestat(\theta,T) \over \fonestat(\theta',T)} \right)
\,.
\ees
The observables $\Phi_1^{\rm rel}$ and $\Phi_2$ are computed in (quenched) QCD, 
with the O($a$)-improved Wilson action as in~\cite{HQET:pap2}.
The light quark mass is fixed to be zero.
While we use different resolutions $L_1/a=20,24,32,40$, the physical 
size of the lattice is kept fixed by determining the bare coupling
such that the SF coupling is~\footnote{We have not yet taken the
error in $g^2(L_1)$ into account. It is expected to be negligible compared to the
other errors.} $\bar g^2(L_1/4)=1.88$~\cite{mbar:pap1,HQET:pap2}.
Using $r_0=0.5\,\fm$ to convert to physical units,
one finds $L_1\approx0.36$ fm.
We take three different values 
$\theta,\theta' = 0,\ 0.5,\ 1$ for the periodicity phases. 
We want to interpolate our data to the b quark mass, 
and since according to~\cite{hqet:pap1}, the value of $L_1 M_{\rm b}^{\rm RGI}$
is around 12, we simulate three different values of $z=L_1M$ : $10.3, 12, 13.2$.
For each of these values 
$\Phi_2(L_1,M)$ is extrapolated to the continuum 
via a linear fit in $(a/L_1)^2$ using only the lattices of resolution
$L_1/a=24,32,40$. We confirmed that adding the coarsest lattice
does not change the values of the continuum limit, as one can also see in
fig.~\ref{phi2}.
We then fit the continuum values - 
see also eq.~(I.5.5) -
to the following form
\bes
\Phi_2(L_1,M)=\mbox{Const}+S\times L_1M\,,
\ees
to find the slope $S=0.61(5)$. 
At this point, 
we note that a source of error originates from the determination~\cite{HQET:pap2}
of the (universal) renormalization factor 
$M_{\rm b}^{\rm RGI}/ {\overline m_{\rm SF}(1/L_0)}$. It amounts to 
a one percent uncertainty on $\Phi_2$.
%
%

\subsection{Effective theory}

Using the static actions denoted by HYP1 and HYP2 in~\cite{stat:actpaper}, 
we compute 
boundary to boundary correlation functions of the QCD 
Schr\"odinger functional, but now in the effective theory and for 
two values of $L$.
The smaller one is again $L=L_1$,
but with the resolutions $L_1/a = 6,8,10,12$, corresponding to $\beta =\ 6.2204,\  
6.476,\  6.635,\  6.775\ $ respectively and the larger one is $L=L_2=2L_1$
with the same values of $a$.
As we did for the relativistic part, for each of these volumes and resolutions,
the simulations are done with three values of the time extent, namely $T = \ L/2,\ L/2-a,\ L/2+a$,
and three values of the periodicity angles.
Following~\cite{hqet:Raiproc},  
we compute $f_1^{\rm stat}$ and $f_1^{\rm kin}$ for these different sets 
of parameters and from those the functions 
$\Gamma_1^{\rm stat}$, $\Gamma_1^{\rm kin}$ and $R_1^{\rm kin}$ as they are defined in eqs.
(I.2.8 - I.2.10) -- at finite values of the lattice spacing.


\section{Step scaling functions}

\begin{figure}[t]
\begin{center}
\leavevmode
\includegraphics[scale=0.3,clip,angle=270]{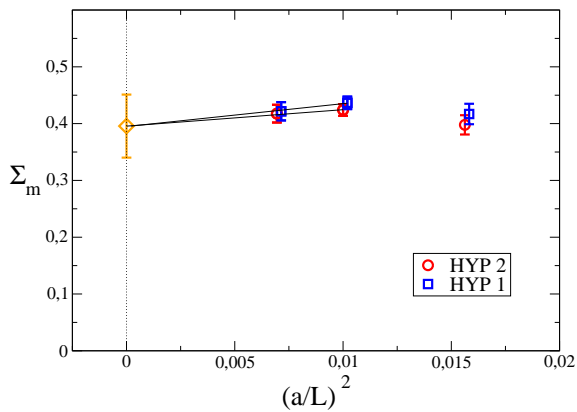}
\end{center}
\vspace{-0.5cm}
\caption[]{\label{sigmam} 
{Continuum extrapolation of the static step scaling function}}
\end{figure}

\begin{figure}[t]
\vspace{-0.4cm}
\begin{center}
\begin{tabular}{cc}
\includegraphics[scale=0.3,clip,angle=270]{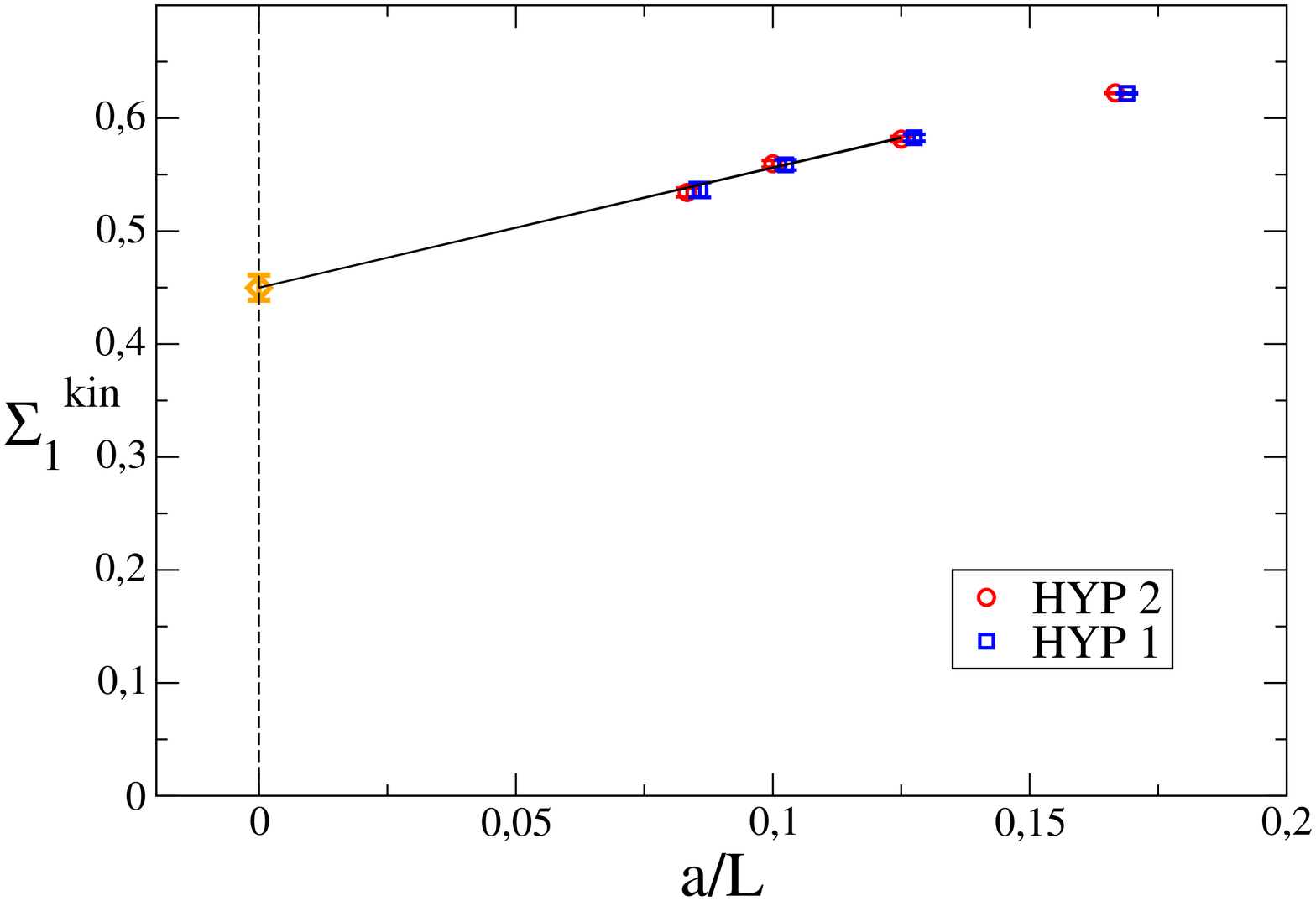} &
\includegraphics[scale=0.3,clip,angle=270]{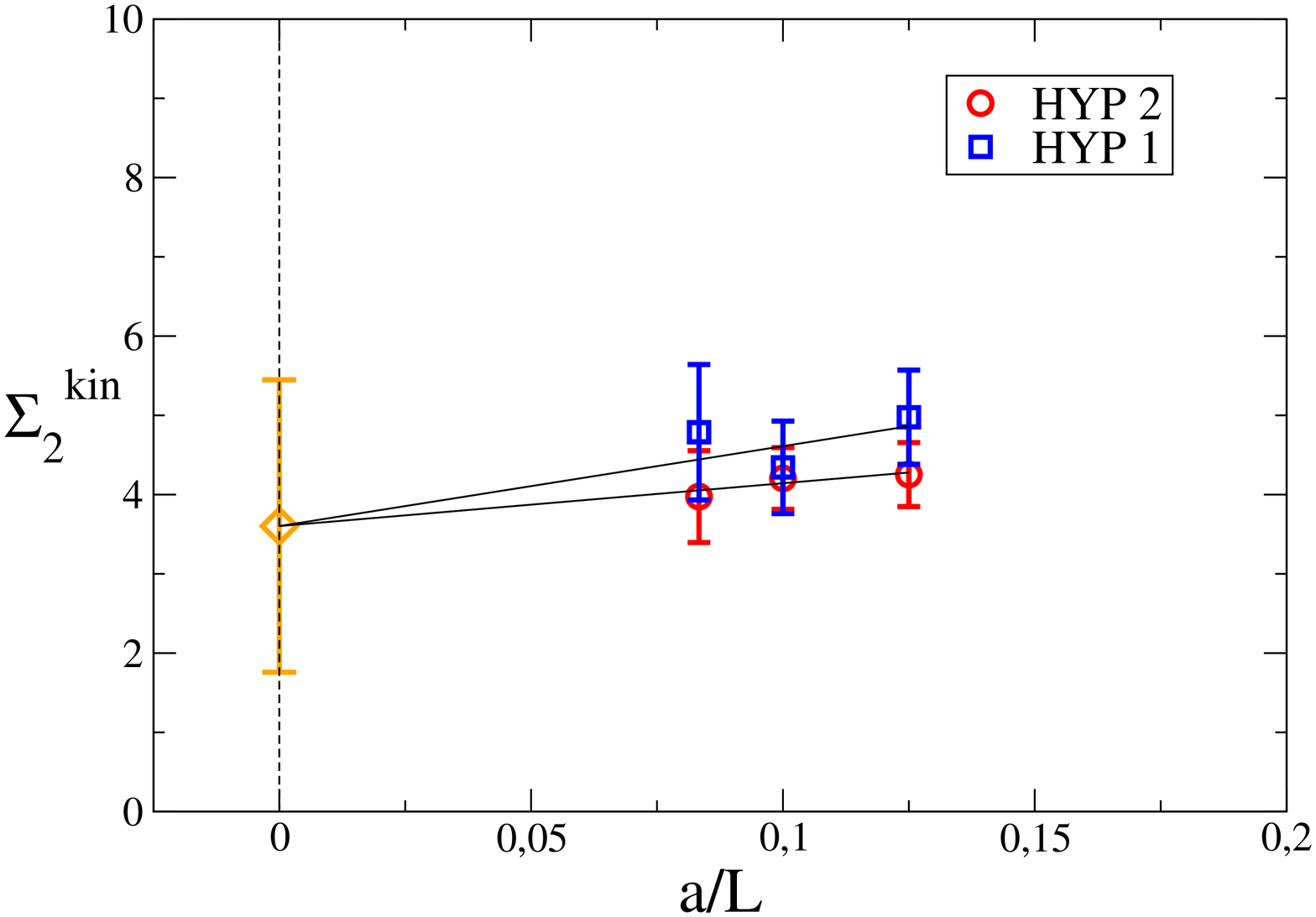} \qquad \\
\end{tabular}
\end{center}
\vspace{-0.5cm}
\caption[]{\label{sigmakin} 
{Continuum extrapolation of the kinetic step scaling functions}}
\end{figure}

Once we have the quantities described in the previous section, 
we compute the step scaling functions $\sigma_m$, 
$\sigma_{1,2}^{\rm kin}$, defined as in eqs (I.3.1 - I.3.3).
At finite lattice spacing, we have 
\bes
\Sigma_m(u,a/L_1) &=& 2L_1 \left[\Gamma_1^{\rm stat}(2L_1,a)-
                                 \Gamma_1^{\rm stat}(L_1,a)\right]_{u=\gbar^2(L_1)}
\ees
%
%
\bes
\Sigma_1^{\rm kin}(u,a/L_1) = \left. {\ratonekin(2L_1,a) \over \ratonekin(L_1,a)} \right|_{u=\gbar^2(L_1)} 
\quad 
\Sigma_2^{\rm kin}(u,a/L_1) = 2L_1 \left. {{\meffkin_1(2L_1,a)-\meffkin_1(L_1,a)}
\over {\ratonekin(L_1,a)}} \right|_{u=\gbar^2(L_1)} \hspace{-0.3cm} . \hspace{0.5cm}
\ees
Their continuum limits 
\bes
\sigma_m(u)=\lim_{a/L_1\rightarrow 0}\Sigma_m(u,a/L_1) 
\qquad
\sigma_{1,2}^{\rm kin}(u)=\lim_{a/L_1\rightarrow 0}\Sigma_{1,2}^{\rm kin}(u,a/L_1) 
\ees
are well defined. Since the static theory is  O($a$)-improved,
$\sigma_m(u)$ is obtained by a linear fit in $(a/L_1)^2$, and we
choose to use only the two finest lattices, corresponding
to $L_1/a=10,12$\,. 
The kinetic step scaling functions $\sigma_{1,2}^{\rm kin}(u)$
are extracted by fitting the data of the three finer lattices 
linearly in $a/L$.
Both for $\Sigma_1^{\rm kin}$ and for $\Sigma_{\rm m}$ 
we could have included one coarser lattice without a significant change 
(apart from smaller errors) as illustrated in figs. 
\ref{sigmam},\ref{sigmakin}.~\footnote{Since the definition of $\Sigma_m$ 
and $\Sigma_2^{\rm kin}$ involves simulations on a lattice of time 
extent $T=L/2-a$, a space extent of $L_1/a=6$ is too small to obtain 
trustworthy numbers. For this reason in the plots of these step scaling
functions, we do not show the data coming from this lattice.}

Since we find well compatible results from the different actions 
after taking the continuum limits, 
we choose a constrained fit in 
the final analysis. 
In that case, the error is computed by a jackknife procedure,
binning the data in such a way that the number of jackknife configurations
is the same for the different sets of data.
We note that a very significant $\theta,\theta'$ dependence is still 
left in the kinetic step scaling functions (I.2.8), nevertheless the 
final physical results will turn out to be independent 
of these kinematical variables as they should (apart from
small $1/m^2$ terms).
In our figures, we made the choice 
$(\theta,\theta')=(0,0.5)$. 
%
%
%
\section{Large volume}

The quantities computed in ``infinite'' volume are the static 
and the kinetic energy (I.4.2), where now the light quark
mass is fixed to $m_\mrm{strange}$. These energies occur in eq. (I.4.1),
through the terms 
$[E^{\rm stat}-\Gamma_1^{\rm stat}(L_2)]$ and 
$[\omega_{\rm kin}({\hat E}^{\rm kin}-\Gamma_1^{\rm kin}(L_2))]$ . 
The infinite and the finite volume part of the static and the kinetic energy
are of course computed at the same value of the lattice spacing (in order to cancel 
the divergences, as we did for the step scaling function).
Our approximation to infinite volume is $L_{\infty}=4 L_1 \simeq 1.5$ fm
(again using $r_0=0.5\,\fm$),
and we use two resolutions $L_{\infty}/a=16,24$,
corresponding to $\beta=6.0219\,,\;6.2885$ respectively.
We obtain a nice signal 
for $\Estat$ due to the HYP1/2 actions~\cite{stat:actpaper}. 
The errors in the effective mass remain below a level of 
few MeV out to 2~fm time separation. Thus the significant excited
state contamination can be removed with confidence and 
$E^{\rm stat}-\Gamma_1^{\rm stat}(L_2) = 612(2)\,\MeV $ and $597(2)\,\MeV$ are
found at $\beta=6.0219\,,\;6.2885$ respectively for the action HYP2. 
From this we quote at present $E^{\rm stat}-\Gamma_1^{\rm stat}(L_2) = 586(11)\MeV $ 
as a continuum number.

However, $[\omega_{\rm kin}({\hat E}^{\rm kin}-\Gamma_1^{\rm kin}(L_2))]$ is 
very small and the errors of the effective ($x_0$-dependent)
${\hat E}^{\rm kin}$ grow rather rapidly with $x_0$. Thus, although we 
will see that this error is not dominant in the end, 
$[\omega_{\rm kin}({\hat E}^{\rm kin}-\Gamma_1^{\rm kin}(L_2))]$
has a large {\em relative} error.

\section{The RGI b quark mass }

\subsection{Static part}

\begin{figure}[t]
\begin{center}
\includegraphics[scale=0.4,clip,angle=270]{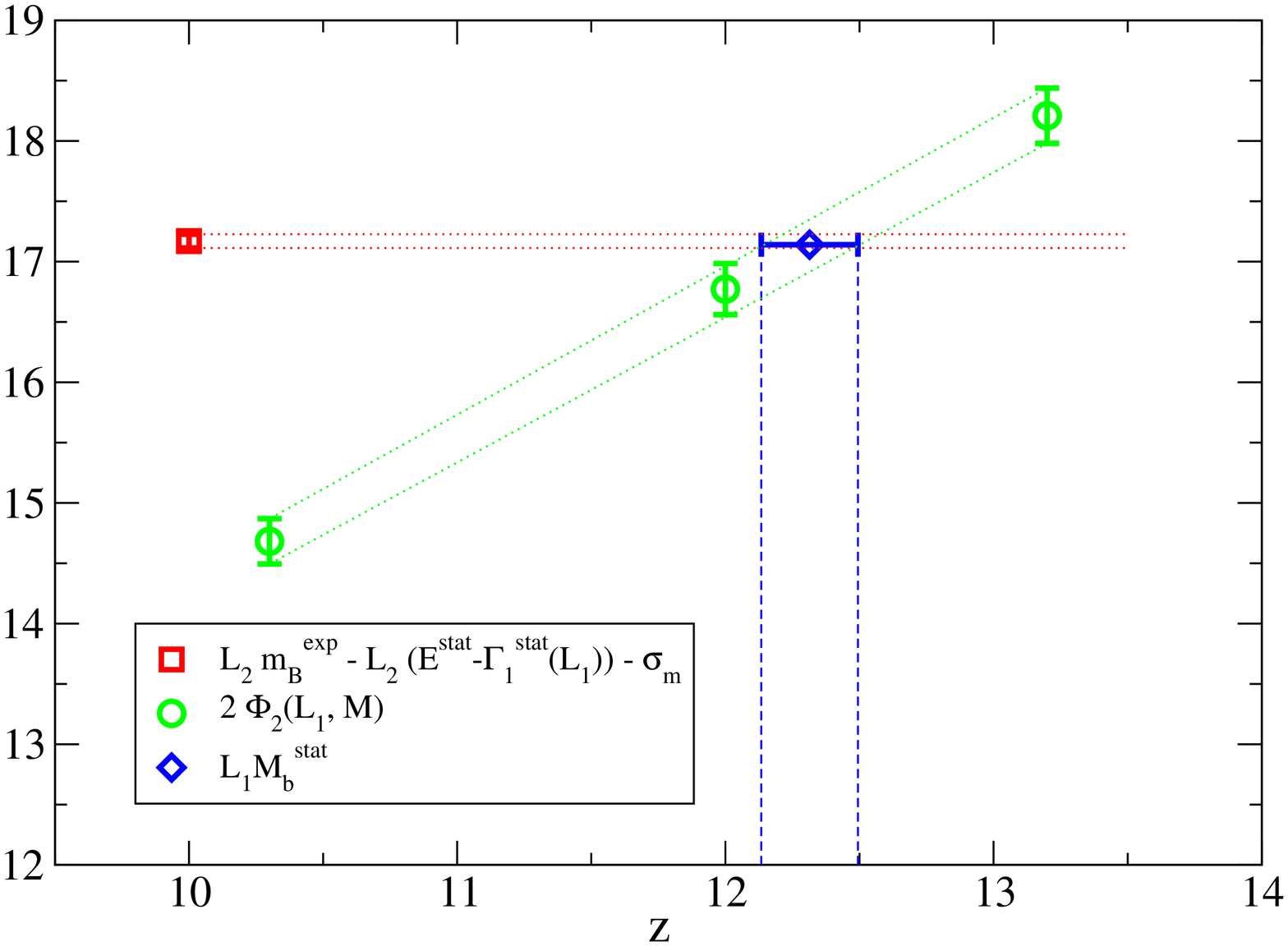}
\end{center}
\vspace{-0.5cm}
\caption[]{\label{Interpol} {Interpolation and solution of eq. (I.5.2)}}
\end{figure}

The value of the RGI quark mass in the static approximation, defined by
eq. (I.5.4), is obtained by a linear interpolation of (I.5.2), where 
we used $m_{\rm B}^{experimental}=5404$ 
MeV as appropriate for a strange quark. 
This interpolation is illustrated in fig \ref{Interpol}.
We obtain (the conversion to physical units is done with $r_0=0.5\,\fm$)
\bes
r_0M_{\rm b}^{\rm stat} = 
\left\{ \begin{array}{ll} 
17.18(25) \mbox{ (HYP2)} \\
17.15(25) \mbox{ (HYP1)}
\end{array} 
\right.
%
%
\qquad
M_{\rm b}^{\rm stat} = 
\left\{ \begin{array}{ll} 
6771(99) \mbox{ MeV (HYP2)} \\
6757(99) \mbox{ MeV (HYP1)}
\end{array}
\right.
\ees

\subsection{$1/m$ corrections}
\label{oneovermsection}
The next to leading order correction of the b quark mass can be separated in two parts,
corresponding to $m_{\rm B}^{\rm (1a)}$, computed entirely from the small volume,  
and $m_{\rm B}^{\rm (1b)}$, given by eq. (I.5.3). 
Following eq. (I.5.6) we have
\bes 
M_{\rm b}^{\rm (1a)} &=& -{1 \over S} \sigmakin_2(\gbar^2(L_1))\Phi_1(L_1,M_{\rm b}^{\rm stat} ) \label{M1a}\\
M_{\rm b}^{\rm (1b)} &=& -{1 \over S} \left [L_2\,(\hat E^\mrm{kin} - \meffkin_1(L_2)) \omegakin\right]\label{M1b} ,
\ees
and we find for the action HYP2
\bes
\nonumber
\begin{array}{||c|c||c|c||}
\hline
\hline
\theta &\theta'& r_0M_{\rm b}^{\rm (1a)} &   r_0M_{\rm b}^{\rm (1b)}\\
\hline
\ \ 0  \ \ & \ \ 0.5 \ \ & \quad -0.08(4) \quad   & \quad  -0.14(11) \quad    \\
\ \ 0.5\ \ & \ \ 1   \ \ & \quad -0.08(4) \quad   & \quad  -0.15(11) \quad    \\
\ \ 1 \ \  & \ \ 0   \ \ & \quad -0.08(4) \quad   & \quad  -0.15(11) \quad    \\
\hline
\hline
\end{array}
\ees
while for HYP1 the numbers are compatible but have larger uncertainties. 
It is now clear that the results are $\theta, \theta'$-independent. In physical units,
we find the scale and scheme independent numbers
%
\bes
M_{\rm b}^{\rm (1a)}=  -30(15) \ \MeV \qquad
M_{\rm b}^{\rm (1b)}=  -56(43) \ \MeV 
\label{Mb1}
\ees
in the quenched approximation.

\section{Conclusions}

Our numerical results show that indeed the $1/m$ corrections 
can be computed with a precision which is -- at present --
better than the one of the leading (static) result: the absolute
errors of leading and next-to-leading terms are in a ratio
of about two to one. This is a first demonstration of the
practicability of the strategy of~\cite{hqet:pap1} beyond the 
leading order. Still, as mentioned
in the previous section, the errors of correlation functions
with the kinetic insertion grow rather rapidly in time
and when decreasing the lattice spacing. The former behavior
is at present not understood.

The size of 
the $1/m$ corrections $M_{\rm b}^{\rm (1a)}$ and $M_{\rm b}^{\rm (1b)}$ 
is small and confirms the validity of the naive counting in terms
of powers of $\Lambda/\mbeauty$. Further numerical confirmation
of this is provided by the close agreement of our result with
the one - also in quenched QCD - of~\cite{deDivitiis:2003iy}, 
$r_0\Mb=17.08(41)$ where the effective theory was {\em not} used.
This also means that $1/m^2$ terms can be neglected
altogether.

On the other hand,  our static result differs (statistically) 
significantly from the static one of~\cite{hqet:pap1}, 
$r_0\M_{\rm b}=16.12(25)(15)$, where the second error is in common 
between that computation and the present one.  
In~\cite{hqet:pap1} the matching was performed using an other 
observable, and at $L=L_0=L_1/2$.
By a comparison to our present NLO result, we see
that this leads to a somewhat larger ($\simeq 5\%$) $1/m$
correction.
This is not at all
unexpected, cf. item 3 in section 6 of [1]. In fact,
an explicit computation of this correction would
be another interesting confirmation of the applicability of the
effective theory.

\noindent
Let us finally translate our numbers to 
the $\msbar$ scheme. With $r_0=0.5\,\fm$ and 
$\Lambda_\msbar^{(0)}=238\,\MeV$~\cite{mbar:pap1}
we find for the b quark mass 
at its own scale
\bes
  m_{\rm b} = m_{\rm b}^{\rm stat} + m_{\rm b}^{(1)} \,,\qquad
  m_{\rm b}^{\rm stat}(m_{\rm b}) = 4.350(64) \,\GeV \,,\quad
  m_{\rm b}^{(1)}(m_{\rm b}) = -0.049(29) \,\GeV \,.
\ees

Despite the employed quenched approximation,
the total result, $m_{\rm b}(m_{\rm b}) = 4.30(7) \, \GeV$, 
is in good agreement with the range quoted in the
particle data book and not far from the precise
value of $m_{\rm b}(m_{\rm b}) = 4.19(5)\,\GeV$
derived in~\cite{mb:steinhkuehn} from the $e^+e^- \to b$ total cross section
and high order perturbation theory. 
Our result is also compatible
with the early computation (quenched and static) of~\cite{mbstat:MaSa} 
$m_{\rm b}(m_{\rm b})=4.41(5)(10)\,\GeV$, where the matching
was done perturbatively at the NLO. 
This last result was updated to 4.30(5)(5)~GeV in~\cite{lat00:lubicz}
with the help of the NNLO result of~\cite{stat:deltamnnlo}.

It seems that on the one hand it is the right time to apply these 
methods both to quantities where larger $1/m$ corrections are expected, 
such as $\Fb$, and to the mass of the b quark in full QCD. 
On the other hand, research should continue to improve the statistical 
errors in particular of the $1/m$ corrections in large volume.
A promising route is to follow~\cite{stat:dublin,pot:stringbreaknf2}.\\

\noindent {\bf Acknowledgement.} 
We thank NIC for allocating computer 
time on the APEmille computers at DESY Zeuthen to this project and the 
APE group for its help.

\bibliography{lat05}           

\begin{thebibliography}{10}

\bibitem{hqet:Raiproc}
ALPHA, M. Della~Morte, N. Garron, M. Papinutto and R. Sommer,
\newblock PoS  (2005) 223, hep-lat/0509084.

\bibitem{PDG04}
Particle Data Group, E. S. et~al.,
\newblock Phys. Lett. B592 (2004) 1.

\bibitem{hqet:pap1}
ALPHA, J. Heitger and R. Sommer,
\newblock JHEP 02 (2004) 022, hep-lat/0310035.

\bibitem{deDivitiis:2003iy}
G.M. de~Divitiis, M. Guagnelli, R. Petronzio, N. Tantalo and F. Palombi,
\newblock Nucl. Phys. B675 (2003) 309, hep-lat/0305018.

\bibitem{Negishi:2005cz}
S. Negishi, H. Matsufuru, T. Onogi and T. Umeda,
\newblock PoS (2005) 208, hep-lat/0510048.

\bibitem{HQET:pap2}
ALPHA, J. Heitger and J. Wennekers,
\newblock JHEP 02 (2004) 064, hep-lat/0312016.

\bibitem{mbar:pap1}
ALPHA, S. Capitani, M. {L\"uscher}, R. Sommer and H. Wittig,
\newblock Nucl. Phys. B544 (1999) 669, hep-lat/9810063.

\bibitem{stat:actpaper}
M. Della~Morte, A. Shindler and R. Sommer,
\newblock JHEP 08 (2005) 051, hep-lat/0506008.

\bibitem{mb:steinhkuehn}
J.H. {K\"uhn} and M. Steinhauser,
\newblock Nucl. Phys. B619 (2001) 588, hep-ph/0109084.

\bibitem{mbstat:MaSa}
G. Martinelli and C.T. Sachrajda,
\newblock Nucl. Phys. B559 (1999) 429, hep-lat/9812001.

\bibitem{lat00:lubicz}
V. Lubicz,
\newblock Nucl. Phys. Proc. Suppl. 94 (2001) 116, hep-lat/0012003.

\bibitem{stat:deltamnnlo}
F. Di~Renzo and L. Scorzato,
\newblock JHEP 02 (2001) 020, hep-lat/0012011.

\bibitem{stat:dublin}
J. Foley et~al.,
\newblock hep-lat/0505023.

\bibitem{pot:stringbreaknf2}
SESAM, G.S. Bali, H. Neff, T. Duessel, T. Lippert and K. Schilling,
\newblock Phys. Rev. D71 (2005) 114513, hep-lat/0505012.

\end{thebibliography}
\bibliographystyle{h-elsevier}




\end{document}